\documentstyle[aasms4,12pt]{article}
\begin{document}

\title{A Model of Metallicity Evolution in the Early Universe}
\author{G. J. Wasserburg\altaffilmark{1} and Y.-Z. Qian\altaffilmark{2}}
\altaffiltext{1}{The Lunatic Asylum,
Division of Geological and Planetary Sciences, California
Institute of Technology, Pasadena, CA 91125.}
\altaffiltext{2}{School of Physics and Astronomy, University of
Minnesota, Minneapolis, MN 55455; qian@physics.umn.edu.}

\begin{abstract}
We apply the phenomenological model used to explain the abundances of
Fe and $r$-process elements in very metal-poor stars in the Galaxy to
[Fe/H] of damped Ly$\alpha$ systems.
It is assumed that
the first stars formed after the Big Bang were very massive and promptly
enriched the interstellar medium to [Fe/H]~$\sim -3$, at which metallicity
formation of normal stars took over. Subsequent Fe enrichment was provided
by Type II supernovae.
The range of [Fe/H] at a given redshift $z$
for damped Ly$\alpha$ systems is explained by the time $t^{\star}$
after the Big Bang at which normal star formation started in
an individual protogalactic system.
The average $t^{\star}$ is
$\approx 80\%$ the age of the universe
for damped Ly$\alpha$
systems at $z\approx 1.5$ to 4.5,
indicating a long delay
between the Big Bang and  
the turn-on of protogalaxies.
It is inferred that a substantial
fraction of the total baryonic matter may not have been aggregated into
protogalaxies where normal star formation had occurred
down to $z\sim 1.5$. The data near $z = 2.2$ suggest that
the rate of turn-on of protogalaxies was initially very low 
and slowly reached a maximum at $\sim 3$~Gyr after the Big Bang.
This may be important in understanding the rate of formation
of quasars.
\end{abstract}

\section{Introduction}
In a recent paper Prochaska \& Wolfe (2000) reported new data on [Fe/H]
for damped Ly$\alpha$ systems at high redshifts.
They also summarized available data
  with redshifts from $z\approx 1.5$ to 4.5
(e.g., Lu et al. 1996; Lu, Sargent, \& Barlow 1997;
Prochaska \& Wolfe 1999).
Their observations are in general accord with
previous studies and expand the database at $z>3$ substantially.
Prochaska \& Wolfe (2000) emphasized that there is a wide spread in
[Fe/H] at a given $z$ and no damped Ly$\alpha$ system has [Fe/H]~$<-2.7$
(see also Lu et al. 1996, 1997). They also pointed out that
there is only a minimal growth of [Fe/H] from $z\approx 4.5$ to 1.5.
In a theoretical study of Fe and $r$-process abundances in very metal-poor
stars in the Galaxy by Wasserburg \& Qian (2000, hereafter WQ),
it was proposed that the first stars formed after the Big Bang were very
massive ($\gtrsim 100\,M_\odot$) and promptly enriched the interstellar
medium (ISM) to [Fe/H]~$\sim -3$, at which metallicity
formation of normal
stars (with masses $\sim 1$--$60\,M_\odot$) took over.
Subsequent Fe enrichment was provided by a subset of Type II
supernovae.
The interpretations of WQ were based on observations of metal-poor stars
in the Galaxy by Gratton \& Sneden (1994), McWilliam et al. (1995),
McWilliam (1998), and Sneden et al. (1996, 1998).
The apparent agreement between the lower bound on [Fe/H] of damped Ly$\alpha$
systems and the critical metallicity [Fe/H]~$\sim -3$ deduced by WQ for
transition from formation of very massive stars to normal stars
suggests that this Fe enrichment model deserves further study.

Here we present a phenomenological model for the range of [Fe/H]
at a given $z$ for damped Ly$\alpha$ systems based on the model
of WQ. It is assumed that formation of normal stars started in a
damped Ly$\alpha$ system (i.e., a protogalactic system turned on)
at a time $t^{\star}$ after the Big Bang.
The assembly of [Fe/H] at a given $z$ is interpreted
as a sampling of $t^{\star}$ ranging from 0 to the age of the universe at
$z$, $t(z)$. The value
$t^{\star}\sim 0$ corresponds to the upper bound on [Fe/H].
Furthermore, the distribution
of $t^{\star}$ at a given $z$ indicates the rate of turn-on of
protogalaxies prior to $t(z)$. The data near $z=2.2$ suggest that this
rate was initially very low and slowly reached a maximum at $\sim 3$~Gyr
after the Big Bang. We describe the Fe
enrichment model in  detail in \S2 and apply it
to explain the data on [Fe/H] of damped Ly$\alpha$ systems
in \S3. Discussion and conclusions are given in \S4.

\section{Fe Enrichment and Abundances in Metal-Poor Stars}
The Fe enrichment model of WQ was developed to explain the relation
between abundances of Fe and $r$-process elements ($r$-elements) in
metal-poor stars in the Galaxy. Meteoritic data on the inventory of
radioactive $^{129}$I and $^{182}$Hf in the early solar system require
at least two distinct Type II supernova sources for the $r$-process
(Wasserburg, Busso, \& Gallino 1996; Qian, Vogel, \& Wasserburg 1998;
Qian \& Wasserburg 2000). These are the high-frequency H events
responsible for heavy $r$-elements with mass numbers $A>130$
(e.g., Ba and Eu) and the low-frequency L events responsible for light
$r$-elements with $A\leq 130$ (e.g., Ag). The recurrence timescales for
the H and L events are $\Delta_{\rm H}\approx 10^7$~yr and
$\Delta_{\rm L}\approx 10^8$~yr, as required by replenishment of
the appropriate radioactive nuclei in a standard mixing mass ($\sim$ the
size of a molecular cloud) for a supernova. Additional evidence in support
of different sources for the heavy and light $r$-elements
has been found by Sneden et al. (2000).

The observed wide dispersion in abundances of the heavy
$r$-elements such as Ba and Eu over a narrow range of
[Fe/H]~$\sim -3$ to $-2.8$ (McWilliam et al. 1995; McWilliam 1998;
Sneden et al. 1996, 1998) led WQ to conclude that
the H events cannot
produce a significant amount of Fe. In contrast, there is
a correlation
between abundances of Fe and the heavy $r$-elements at
[Fe/H]~$\gtrsim -2.5$ (Gratton \& Sneden 1994; see also
McWilliam et al. 1995).
As Type Ia supernovae would occur
only at metallicities much higher than [Fe/H]~$=-2.5$, Fe enrichment
of very metal-poor stars with
[Fe/H]~$\gtrsim -2.5$ must be provided by the L events. Consequently,
the L events are responsible for the part of Fe contributed by
Type II supernovae in general. There are $\approx 100$ L events
during the time of $\approx 10^{10}\ {\rm yr}$ prior to
solar system formation.  To 
provide $\approx 1/3$ of the solar Fe inventory by these events,
each L event must enrich a standard mixing
mass with [Fe/H]$_{\rm L}\approx -2.5$.

The near absence of the heavy $r$-elements in stars with
[Fe/H]~$\sim -4$ to $-3$ (McWilliam et al. 1995; McWilliam 1998)
and the sharp increase in the abundances of such elements
at [Fe/H] $\sim -3$ to
$-2.8$ led WQ to conclude
that a source other than Type II supernovae must
exist to produce Fe (along with other
elements such as C, N, O, Mg, and Si)
at [Fe/H]~$\lesssim -3$.
This source was attributed by WQ to very
massive stars (with masses $\gtrsim 100\,M_\odot$) that first formed
after the Big Bang. They further argued that formation of normal stars
could not occur until [Fe/H]~$\sim -3$ was reached. Presumably, this
critical metallicity corresponds to conditions in the ISM that permit
sufficient cooling and fragmentation to occur in collapsing gas clouds.
A recent study by Bromm, Coppi, \& Larson (1999) suggests that
the very first stars were  rather massive.
The products of very massive stars  formed from Big Bang debris
have been discussed earlier
by Ezer \& Cameron (1971). However, nucleosynthesis
in such stars remains to be tested
with adequate stellar models.

\section{[Fe/H] of Damped Ly$\alpha$ Systems}
To discuss [Fe/H] of a damped Ly$\alpha$ system at a given $z$,
we assume the following history for its evolution:
(1) at time $t_1$ after the Big Bang, matter consisting of Big Bang
debris was isolated to form a system;
(2) prompt enrichment by the first very massive stars ended at time
$t_2$, resulting in [Fe/H]~$\sim -3$ in the ISM; and (3) formation
of normal stars began at time $t^{\star}$, with the L events providing
further Fe enrichment to an average ISM at regular intervals.
The above times are related as $t_1<t_2<t^{\star}$. As
$t^{\star}\leq t(z)$, it is necessary to define the value
of $t(z)$ that is used. We take the redshift $z$ to
correspond to a time
\begin{equation}
t(z)\approx{2\over 3}H_0^{-1}\Omega_{\rm m}^{-1/2}(1+z)^{-3/2}
\label{tz}
\end{equation}
after the Big Bang. We take the Hubble
constant $H_0=65\ {\rm km\ s}^{-1}\ {\rm Mpc}^{-1}$ and the matter
contribution to the critical density $\Omega_{\rm m}=0.3$
[equation (\ref{tz}) gives essentially the
exact result for $t(z)$ at $z\geq 1.5$ 
for a flat universe with $\Omega_{\rm m}=0.3$
and a cosmological constant]. By our assumption, for
$t(z)\geq t_2$
the metallicity of a damped Ly$\alpha$ system is at least
[Fe/H]~$\sim -3$. Prochaska \& Wolfe (2000) reported that no
system at $z\approx 1.5$ to 4.5 has [Fe/H]~$<-2.7$. This means that
the time required to provide prompt Fe enrichment must be less than
$t(z\approx 4.5)\approx 1.4$~Gyr.
The lower bound on [Fe/H] due to prompt enrichment is shown as
a band between $-3$ and $-2.8$ in Figure 1 together with the data
summarized in Prochaska \& Wolfe (2000).

In general, for $t(z)\geq t^{\star}$ the metallicity of a specific
damped Ly$\alpha$
system at $z$ is given by
\begin{equation}
({\rm Fe/H})=({\rm Fe/H})_{\rm p}+({\rm Fe/H})_{\rm L}
{t(z)-t^{\star}\over\Delta_{\rm L}},
\label{fe}
\end{equation}
where the number ratio in round brackets (Fe/H) is related
to the standard square bracket
notation by [Fe/H]~$=\log({\rm Fe/H})-\log({\rm Fe/H})_\odot$,
with (Fe/H)$_{\rm p}$ corresponding to the prompt
enrichment and (Fe/H)$_{\rm L}$ to the contribution from
a single L event to a standard reference mass of hydrogen.
We take [Fe/H]$_{\rm p}=-2.8$, [Fe/H]$_{\rm L}=-2.5$,
and $\Delta_{\rm L}=10^8$~yr,
the same parameters used by WQ.
As recognized by
Lu et al. (1996), the scatter in [Fe/H] at a given $z$
for damped Ly$\alpha$ systems
might result from their different formation histories.
Equation (\ref{fe}) explicitly states that the range of [Fe/H] at $z$
is caused by the different start times $t^{\star}$ for normal star
formation in individual systems (see Figure 2a). The values for [Fe/H]
corresponding to $t^{\star}=1$ and 1.5~Gyr are shown in Figure 1.
Damped Ly$\alpha$ systems that just turned on at $t^{\star}\approx t(z)$
would have [Fe/H]~$\approx$~[Fe/H]$_{\rm p}$.
An upper bound on [Fe/H] exists as damped Ly$\alpha$ systems that turned
on at $t^{\star} \sim 0$ would have the longest history of normal star
formation, and hence, the highest [Fe/H]. This bound is
insensitive to the choice of [Fe/H]$_{\rm p}$.
It can be seen from Figure 1 that almost all the data lie below the
upper bound (usually well below this bound).
Even the three exceptions are close to this bound (see \S4).

At a fixed $z$, equation (\ref{fe}) with equation (\ref{tz}) can be used
to determine the start
time $t^{\star}$ for normal star formation in a damped Ly$\alpha$ system
from its [Fe/H]. In turn, a histogram of
the number of systems at $z$ within
a given [Fe/H] interval determines the probability
$p\left(t^{\star},t(z)\right)dt^{\star}$ for normal star formation to
start in the
interval between $t^{\star}$ and $t^{\star}+dt^{\star}$ after the Big Bang.
Knowing the probability distribution $p \left(t^{\star},t(z)\right)$ over
$0<t^{\star}\leq t(z)$ at $z$, we expect that the probability distribution
at $z'>z$ can be obtained by discarding the part of
$p\left(t^\star,t(z)\right)$
for $t^\star>t(z')$ and renormalizing the remaining part over
$0<t^\star\leq t(z')$ (see Figure 2b).
For the case of a simple power-law distribution
$p\left(t^\star,t(z)\right)=[(\alpha+1)/t(z)][t^\star/t(z)]^\alpha$.
In this case, the average start time for normal star formation
in damped Ly$\alpha$ systems at $z$ is
$\langle t^\star(z)\rangle=[(\alpha+1)/(\alpha+2)]t(z)$.

The data in Figure 1 show a relatively high concentration in the
interval $2.0\leq z\leq 2.4$. The average start time in
  this interval is $\langle t^\star(z=2.2)\rangle\approx 2.5$~Gyr.
The frequency of occurrences of
$t^{\star}$ calculated from the data in this interval is shown
as a histogram in Figure 3.
It can be seen that the frequency of occurrences is low for
$t^{\star} \sim 0$ (close to the Big Bang) and slowly increases to a maximum
at $t^{\star}\sim 3$~Gyr.
Assuming a power-law distribution for $t^\star$, we obtain
$\alpha\approx 3$, for which $\langle t^\star(z)\rangle\approx 0.8t(z)$
(see the corresponding curve for [Fe/H] shown in Figure 1).
Values of $t^\star$ for
all the data  are shown in Figure 4. The clustering of
$t^\star$ close to $t(z)$ indicates that typically there is
a long delay between the Big Bang and the start time $t^\star$
for normal star formation in damped Ly$\alpha$ systems.

\section{Discussion and Conclusions}
We consider  the dominant cause of dispersion of [Fe/H]
at a given $z$ for damped Ly$\alpha$ systems to be the variation
in the start time $(t^{\star})$ after the Big Bang
for normal star formation in different protogalaxies. The bounds
on [Fe/H] from our model appear to closely define the
observed ranges. The average $t^{\star}$
for damped Ly$\alpha$ systems at $z$ is
$\langle t^{\star}(z)\rangle\approx 0.8t(z)$.
The rate of turn-on of protogalaxies was initially
very low and slowly increased to
a maximum at $\sim 3$~Gyr after the Big Bang.
It is not possible to identify a
turnover in this rate without more data  at $z<2$.
We suggest that the approach outlined here is a
method for dating the start time for normal star formation
in protogalaxies. As the formation of quasars is closely related to
star formation, we consider that the histogram shown in Figure 3
may offer some insights into the rate of formation of quasars.
For example, if the rate of quasar formation $R_{\rm Q}$ is
proportional to the turn-on rate of protogalaxies, Figure 3
suggests that $R_{\rm Q}\propto (t^\star)^3\propto (1+z)^{-4.5}$
at $z\geq 2$. This gives a decrease by a factor of 2.7 in $R_{\rm Q}$
from $z=3$ to 4, consistent with the drop in quasar comoving
space density at these redshifts (Schmidt, Schneider, \& Gunn 1995).

The three data points that lie above the
upper bound in Figure 1 could be
explained if the enrichment rate
$\beta_{\rm L}\equiv{\rm (Fe/H)}_{\rm L}/\Delta_{\rm L}$
of L events were increased by a factor $\sim 2$.
It is possible that $\beta_{\rm L}$ has a
spread of a factor of 2 or 3.
Alternatively, we can consider the enrichment rate as
a steep function of time, with $\beta_{\rm L}(t)$
starting quite high and then settling down to the value
proposed by WQ. In this case (Fe/H)
$=$~(Fe/H)$_{\rm p}+\int_{t^{\star}}^{t(z)}\beta_{\rm L}(t')dt'$
and the straight
line evolution in Figure 2a would be replaced by a curve.
We note that
Fe enrichment by Type Ia supernovae would not be significant for
most damped Ly$\alpha$ systems as such enrichment appears
to be significant
only at [Fe/H]~$>-1$ in the Galaxy (Timmes, Woosley, \& Weaver 1995).

A more difficult matter is the timescale for condensing
protogalactic globs during the expansion of the universe
and the timing
sequence outlined above. The range in $t^{\star}$
required is large and
indicates that the time required to
form protogalaxies and condense
most of the baryonic matter into stars
is comparable to the age of the universe at $z\sim 2$
($\sim 3.5$ Gyr after the
Big Bang) or possibly even longer. This is in conflict with ab initio
models that report almost complete condensation of dark matter
and possibly cloud or protogalaxy formation at $z\sim 10$--20
(e.g., Kamionkowski, Spergel, \& Sugiyama 1994).
The occurrence of [Fe/H]~$\approx -2.6$ at $z\approx 2$
implies that there are regions where formation of normal stars
did not begin until $\approx 3.5$ Gyr after the Big Bang.
The occurrence of [Fe/H]~$\approx -2.6$ to $-2.4$
in the range of $z\approx 2.0$ to 4.2 indicates that
a large fraction of the baryonic matter has not been condensed
into protogalaxies and stars over
the corresponding extended time range.
If damped Ly$\alpha$ systems
are random samples of the original medium,
this indicates that the reservoir of original uncondensed
baryonic material may not have been seriously diminished over
$\sim 3.5$~Gyr.
Indeed, it is possible that the bulk of this baryonic matter is
dispersed and has not condensed today. Measurements
of damped Ly$\alpha$ systems
at lower redshifts (but at times before Type Ia
supernovae contribute Fe)
would provide a test.
We have no means of establishing
a priori whether the intrinsic rate of
protogalaxy formation decreases
with time (possibly due to decrease in global density) without
appreciably depleting the reservoir of baryonic matter.

A variant of the above scenario is possible
which would not be in conflict with the models that
suggest almost complete condensation of baryonic matter by $z \sim 10$.
The modified scenario would be that most baryonic matter
in the potential well created by
non-baryonic dark matter is rapidly
collected into Ur-protogalaxies, that formation of very massive
stars from Big Bang debris is rapid [$t_2\ll t(z)$],
and that the explosion of these very massive stars
destroys the Ur-protogalaxies and redistributes
matter into the general medium providing
a uniform source of Fe (and other elements
such as C, N, O, Mg, and Si).
The material in this second generation medium
is then the source
for much slower aggregation and  formation
of protogalaxies and stars.
The explosion of very massive stars
is likely to be very energetic. The potential wells
formed by dark matter are considered to have a typical
escape velocity of $\sim 400$~km~s$^{-1}$. The velocity of debris
ejected from  explosion of very massive stars is almost certainly
much larger than this value. The inter-protogalactic medium would then
be supplied with hot matter containing Fe, Mg, O, Si, and C.
A hint in favor of this is found in the recent
detection of O VI quasar absorption systems at low redshifts
(Tripp, Savage, \& Jenkins 2000). 
Knowledge of the explosion dynamics and nucleosynthetic products of
very massive stars formed from Big Bang debris is fundamental to further
progress.

\acknowledgments
We want to thank Roger Blandford and Marc
Kamionkowski for their ongoing interest in this work
and for generously giving their time and providing,
as usual, deep cosmic insights and education. Sir Maarten Schmidt
has been kind enough to expose us to the details of quasar distribution.
This work was supported in part by the Department of Energy under grant
DE-FG02-87ER40328 to Y.-Z. Q. and by NASA under grant
NAG5-4083 to G. J. W., Caltech Division Contribution No. 8728(1059).

\clearpage

\figcaption{A summary of the available data on [Fe/H]
of damped Ly$\alpha$ systems at $z\approx 1.5$ to 4.5
(asterisks: Lu et al. 1996, 1997; squares: Prochaska \& Wolfe 1999, 2000).
The curve labeled $t^{\star}=0$ is the upper bound 
from equation (\protect\ref{fe}) with [Fe/H]$_{\rm p}=-2.8$,
[Fe/H]$_{\rm L}=-2.5$, and $\Delta_{\rm L}=10^8$~yr.  
The curve labeled
$\langle t^\star(z)\rangle=0.8t(z)$ appears to represent the average
trend of the data. Curves for [Fe/H] corresponding to fixed values of
$t^{\star}=1$ and 1.5~Gyr are also shown.}

\figcaption{(a) Schematic diagram of (Fe/H) as a function of
$t^{\star}$ at a fixed $z$.
The baseline (Fe/H)$_{\rm p}$
is the prompt enrichment value.
Normal star formation began in a protogalactic system at time
$t^{\star}$ after the Big Bang. After reaching
(Fe/H)$_{\rm{p}}$, a particular mass of matter must wait until
$t^{\star}$ to start the increase in (Fe/H) by 
contributions from Type II supernova L events.
For $t^{\star} = 0$,
there is a maximum increase in (Fe/H) (trajectory A).
As $t^{\star}$ approaches $t(z)$, the growth is proportionally
smaller (trajectories B and C). (b) Schematic histogram for the
distribution of $t^{\star}$ 
in a set of damped Ly$\alpha$ systems at
a given $z$. The distribution for $z'> z$ can be obtained
by discarding the part of the distribution for $z$ at $t^{\star}>t(z')$.}

\figcaption{Histogram of $t^{\star}$ for data between
$2.0\leq z \leq 2.4$ in Figure 1
showing a peak at $t^{\star}\approx 2.5$~Gyr
when $t(z=2.2)\approx 3.2$~Gyr. This shows that the rate
of turn-on of protogalaxies starts from a low value close to the Big Bang
and increases until at least $z\approx 2$.}

\figcaption{Values of $t^{\star}$ calculated from
equation (\protect\ref{fe}) with [Fe/H]$_{\rm p}=-2.8$,
[Fe/H]$_{\rm L}=-2.5$, and $\Delta_{\rm L}=10^8$~yr for all
data in Figure 1. Note the clustering of
$t^{\star}$ close to $t(z)$ over the range of $z$. 
Three data points indicated by downward arrows
require $t^\star<0$.}
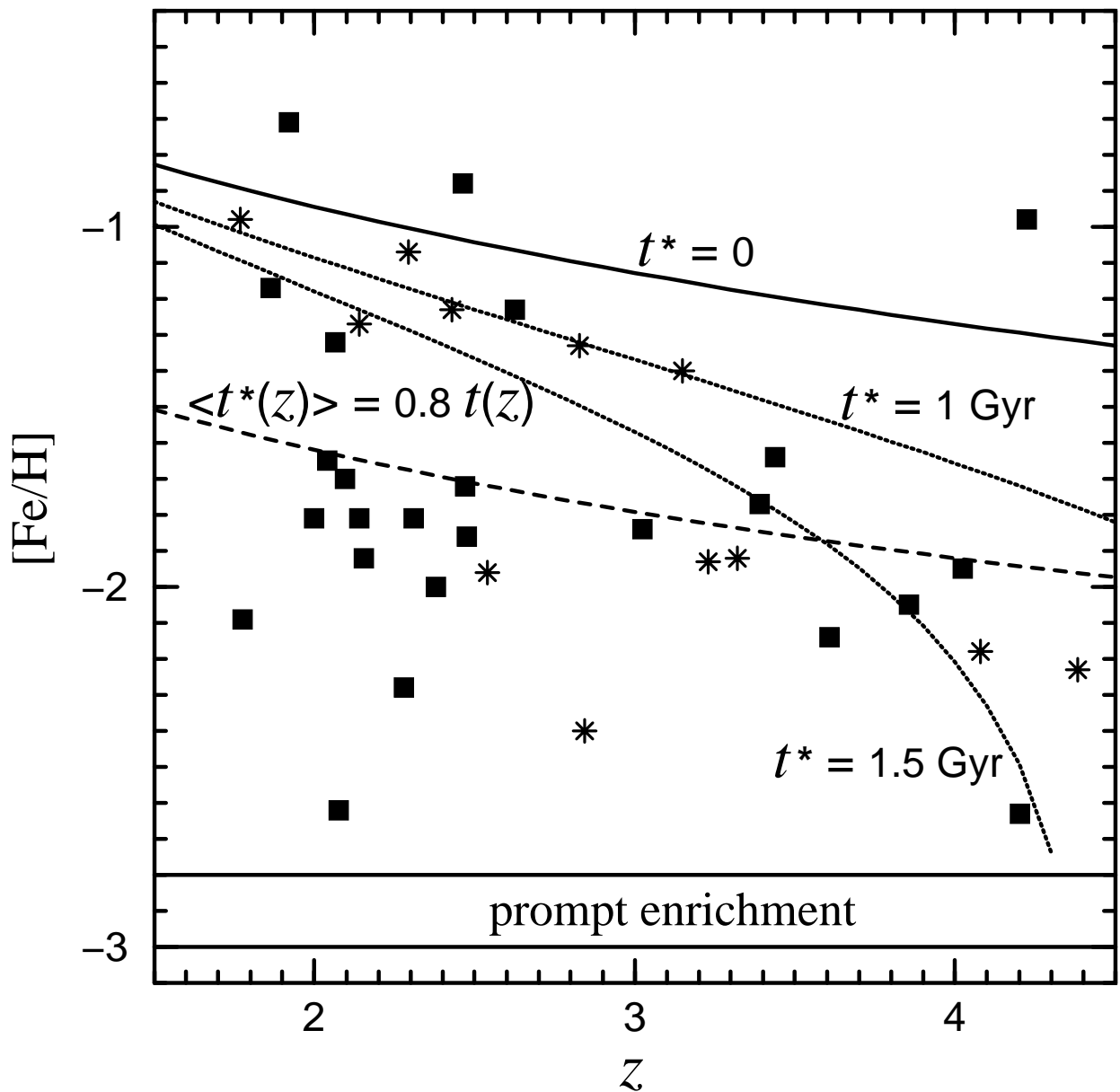
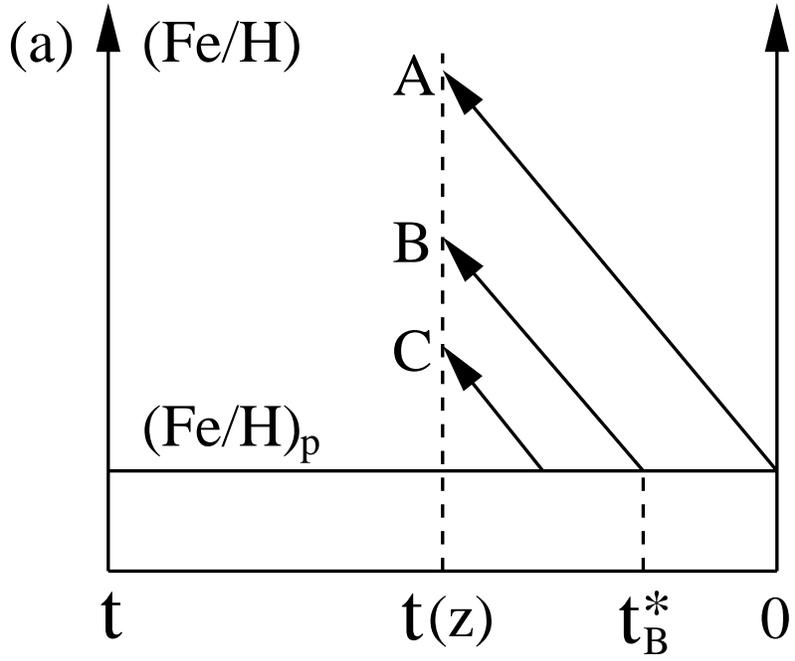
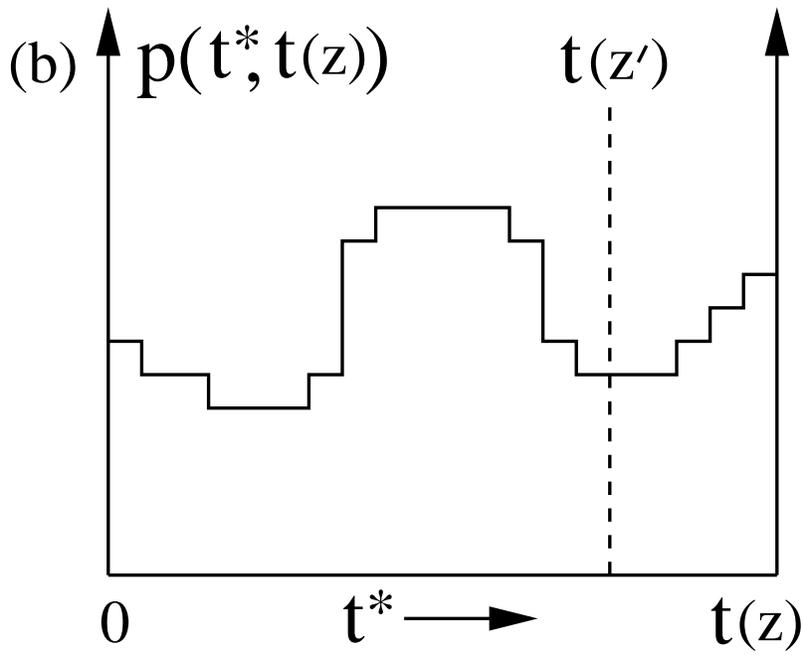
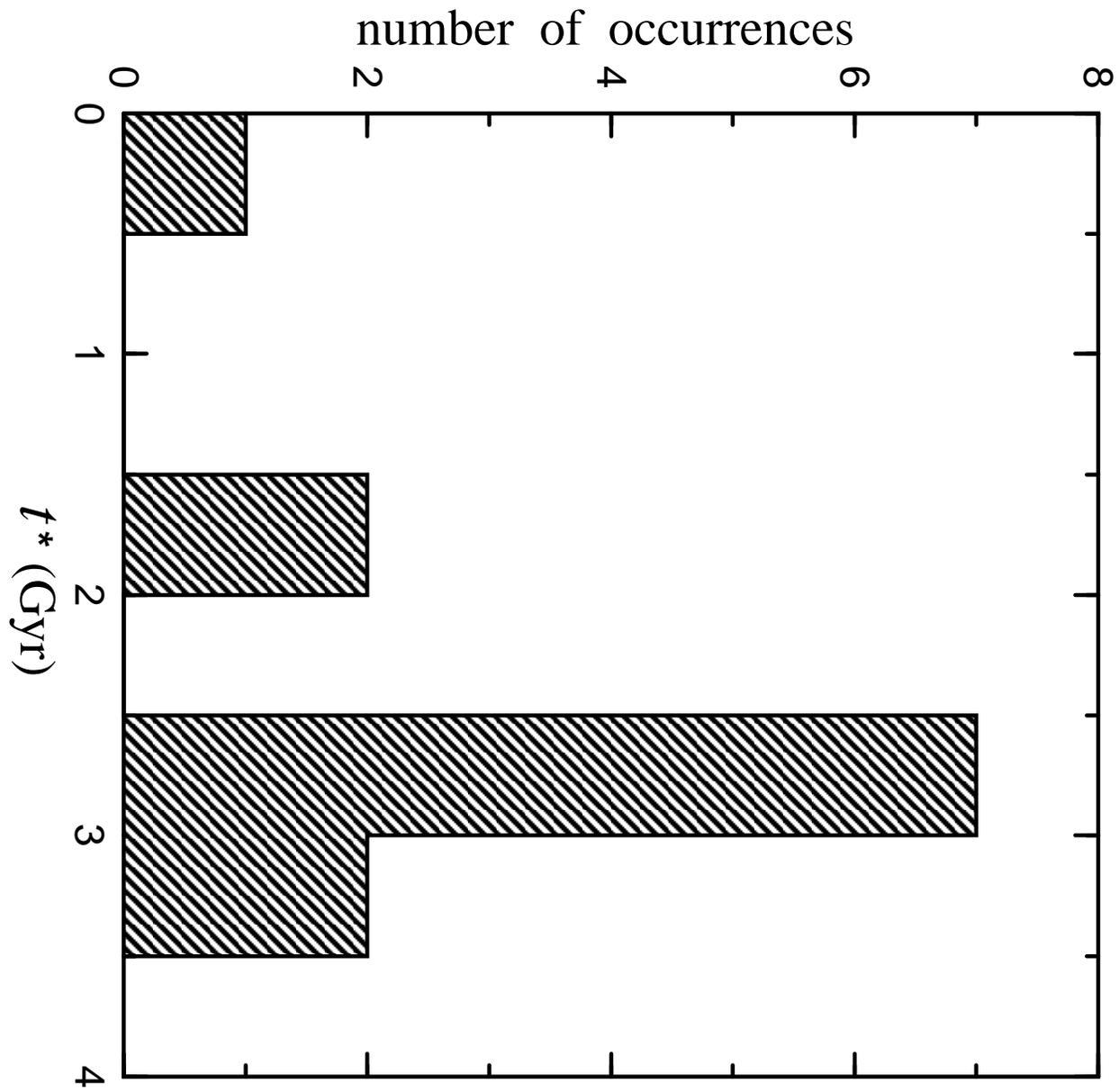
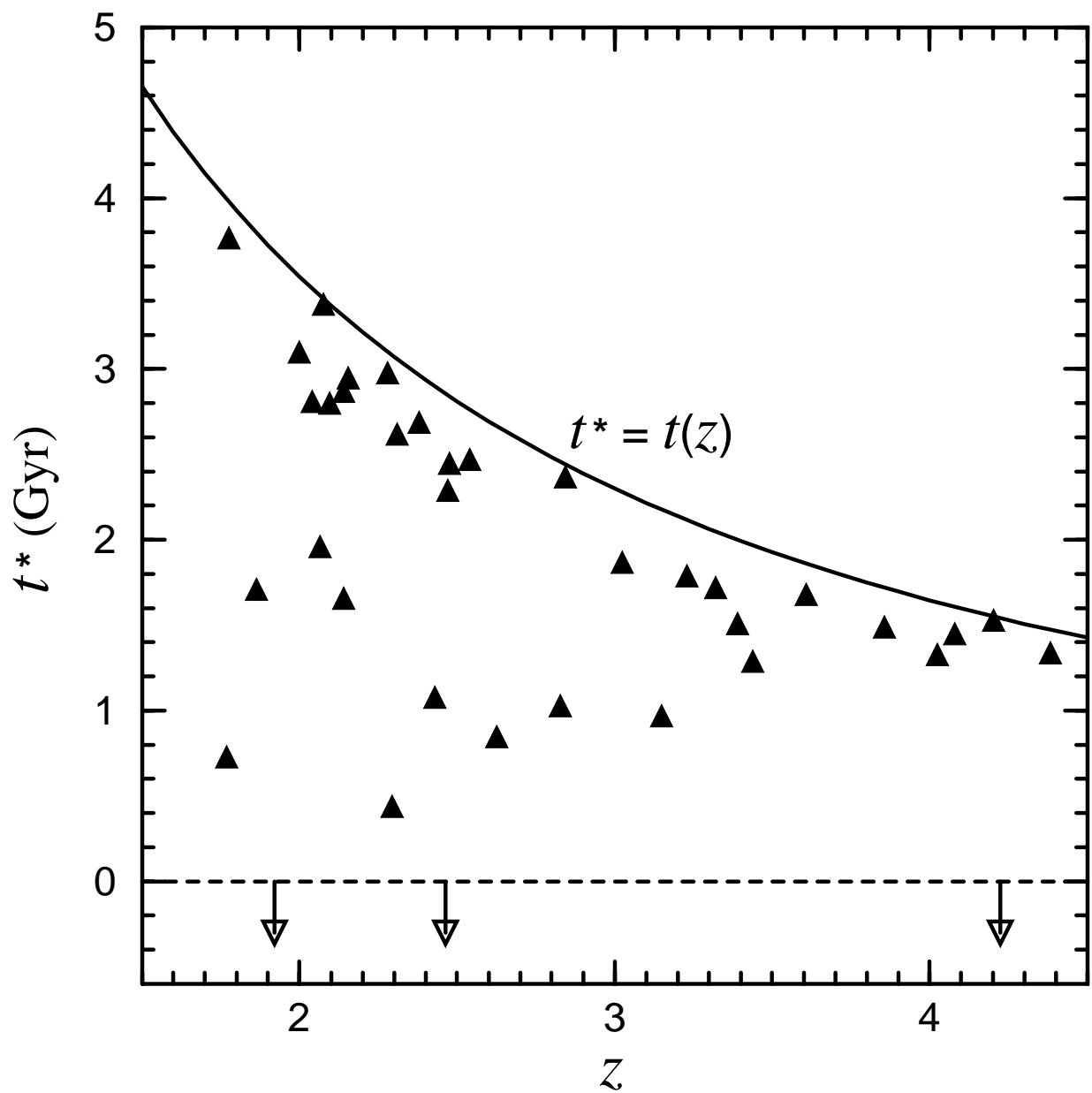
\end{document}